\documentclass{ws-procs9x6}

\begin{document}

\title{SIDIS Asymmetries in Quark-Diquark Model}

\author{Aram KOTZINIAN$^*$}

\address{CEA-Saclay, IRFU/Service de Physique Nucl\'{e}aire,
91191 Gif-sur-Yvette, France\\
On leave in absence from YerPhI, Yerevan, Armenia and JINR, Dubna, Russia
\\
$^*$E-mail: aram.kotzinian@cern.ch}

\begin{abstract}
Some properties of intrinsic transverse momentum dependent nucleon distribution functions are considered in the simple quark-diquark model. The transverse target polarization dependent asymmetries for SIDIS are calculated and compared with recent results of COMPASS. The model describes well the measured asymmetries. Generalization of quark-diquark model for Sivers function is discussed.
\end{abstract}


\bodymatter

\section{Introduction}\label{sec:intro}

The production of one unpolarized hadron in polarized lepton-nucleon DIS can be described using eighteen structure functions~\cite{AK95,B06}\/. Eight of them are describing azimuthal asymmetries proportional to target transverse polarization. Within factorized QCD parton model, these structure functions can be expressed as convolutions of transverse momentum dependent (TMD) quark distribution and fragmentation functions (DFs and FFs). To evaluate these structure functions only six T-even twist-two DFs and two twist-two FFs together with  ${\cal O} (1/Q)$ kinematic corrections were taken into account in~\cite{AK95}\/. The full expressions for the structure functions in terms of convolutions of twist-two and twist-three DFs and FFs are given in~\cite{B06}\/.

Since the TMD DFs and FFs are nonperturbative objects and cannot be calculated from first principles, the standard way to study them is to use some parameterizations and then, using expressions for the cross-sections and asymmetries, fit the parameters to existing data~\cite{Prok08, Bogl08}\/. Other approach --- develop nonperturbative models for nucleon structure and quark fragmentation, see for example~\cite{JMR97, Bacc03, Gamb07, Radi08}\/.

Here I'll discuss some properties of DFs calculated in a simple quark-diquark model~\cite{JMR97} and present the comparison of results for target transverse spin azimuthal asymmetries with COMPASS preliminary data~\cite{AK07,BP07}\/.

We use the model of Ref.~\cite{JMR97} with a modified formfactor for the proton-quark-diquark vertex. Namely, instead of power-like dependence on quark virtuality, we use the Gaussian dependence:
\begin{equation}\label{ff}
    {1 \over |k^2-\Lambda^2|^\alpha} \Rightarrow \exp\left(k^2/2\Lambda^2\right).
\end{equation}
With this choice of formfactor the six leading order (LO) T-even DFs for the scalar ($R=S$) and axial-vector ($R=A$) diquark look like
\begin{eqnarray}\label{6dfs}
  f_1^R(x,{\bf k}_T^2) &=& f_0^R(x)\left((xM+m)^2+{\bf k}_T^2\right)G({\bf k}_T^2,x), \nonumber \\
  g_1^R(x,{\bf k}_T^2) &=& a_R\,f_0^R(x)\left((xM+m)^2-{\bf k}_T^2\right)G({\bf k}_T^2,x), \nonumber \\
  h_1^R(x,{\bf k}_T^2) &=& a_R\,f_0^R(x)(xM+m)^2G({\bf k}_T^2,x), \nonumber \\
  h_{1T}^{\perp R}(x,{\bf k}_T^2) &=& -2a_R\,f_0^R(x)M^2G({\bf k}_T^2,x),  \\
  g_{1T}^{\perp R}(x,{\bf k}_T^2) &=& 2a_R\,f_0^R(x)M(xM+m)^2G({\bf k}_T^2,x), \nonumber \\
  h_{1L}^{\perp R}(x,{\bf k}_T^2) &=& -g_{1T}^{\perp R}(x,{\bf k}_T^2), \nonumber
\end{eqnarray}
where
\begin{eqnarray}\label{gaus}
    &&G({\bf k}_T^2,x)={1 \over \pi \mu^2(x)}\exp\left({\bf k}_T^2/\mu^2(x)\right), \nonumber \\
    &&f_0^R(x)=N_R\exp\left({x(1-x)M^2-xM_R^2 \over \mu^2(x)}\right),\\
    &&\mu^2(x)=\Lambda^2(1-x), \nonumber
\end{eqnarray}
$a_S=1$ and $a_A=-1/3$. The DFs for {\it u} ({\it d}) quarks in the proton are given by \begin{eqnarray}\label{udfs}
f_1^u(x,{\bf k}_T^2)&=&{3 \over 2}f_1^S(x,{\bf k}_T^2)+{1 \over 2}f_1^A(x,{\bf k}_T^2), \nonumber \\
f_1^d(x,{\bf k}_T^2)&=&f_1^A (x,{\bf k}_T^2)
\end{eqnarray}
and similarly for other DFs. Since the model is describing only the valence quarks, the normalization factors $N_S$ and $N_A$ are obtained by requiring that  $f_1^S$ and $f_1^A$ are normalized to unity upon integration over $x$ and ${\bf k}_T^2$. In expressions (~\ref{ff}--\ref{udfs}), $M$ is the proton mass and $m=0.36$ GeV/c, $M_S=0.6$ GeV/c, $M_A=0.8$ GeV/c and  $\Lambda=0.5$ GeV/c are free parameters of the model suggested in~\cite{JMR97}\/.

Let us stress here that since the model is based on tree diagrams, the T-odd Sivers, $f_{1T}^\perp(x,{\bf k}_T^2)$, and Boer-Mulders, $h_{1}^\perp(x,{\bf k}_T^2)$, functions are equal to zero.

\section{Helicity distribution function}\label{sec:hel-df}

In many phenomenological applications for TMD DFs the $x$-$k_T$ factorized form: $f_1^q(x,{\bf k}_T^2)=f_1^q(x)\exp\left({\bf k}_T^2/\mu_1^2\right)/\pi \mu_1^2$, $g_1^q(x,{\bf k}_T^2)=g_1^q(x)\exp\left({\bf k}_T^2/\mu_2^2\right)/\pi \mu_2^2$ with {\it x}-independent width of transverse momentum distribution $(\mu_{1,2}^2=Const)$ is used. With this choice the quark longitudinal polarization, $P_q(x,{\bf k}_T^2)=g_1^u(x,{\bf k}_T^2)/f_1^u(x,{\bf k}_T^2)$, also has a $x$-$k_T$ factorized form. Thus, for different fixed transverse momenta, the quark polarization has similar (re-scaled depending on these transverse momenta factors) x-dependence.

\begin{figure}
\begin{center}
\psfig{file=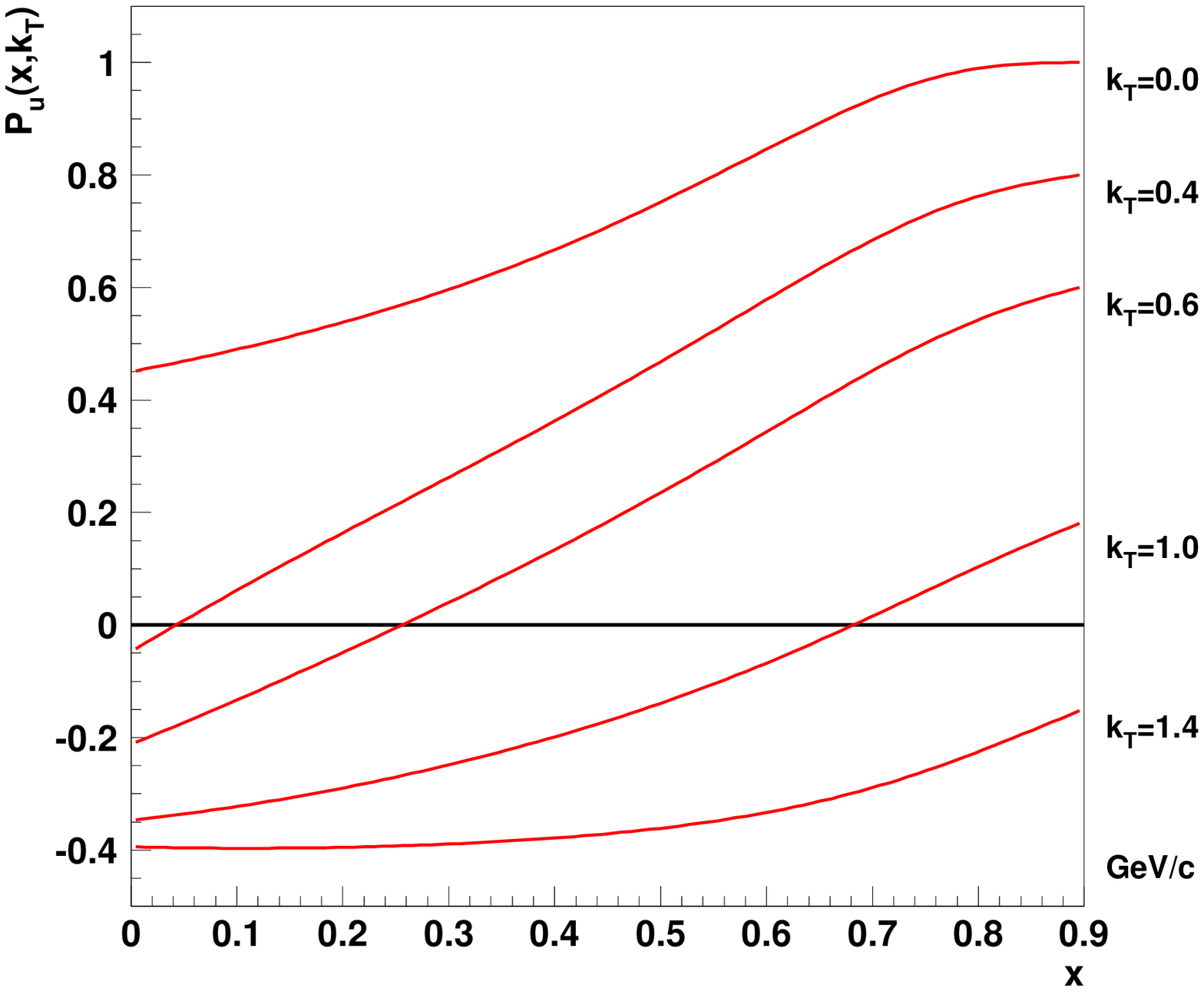,width=5.6cm}
\psfig{file=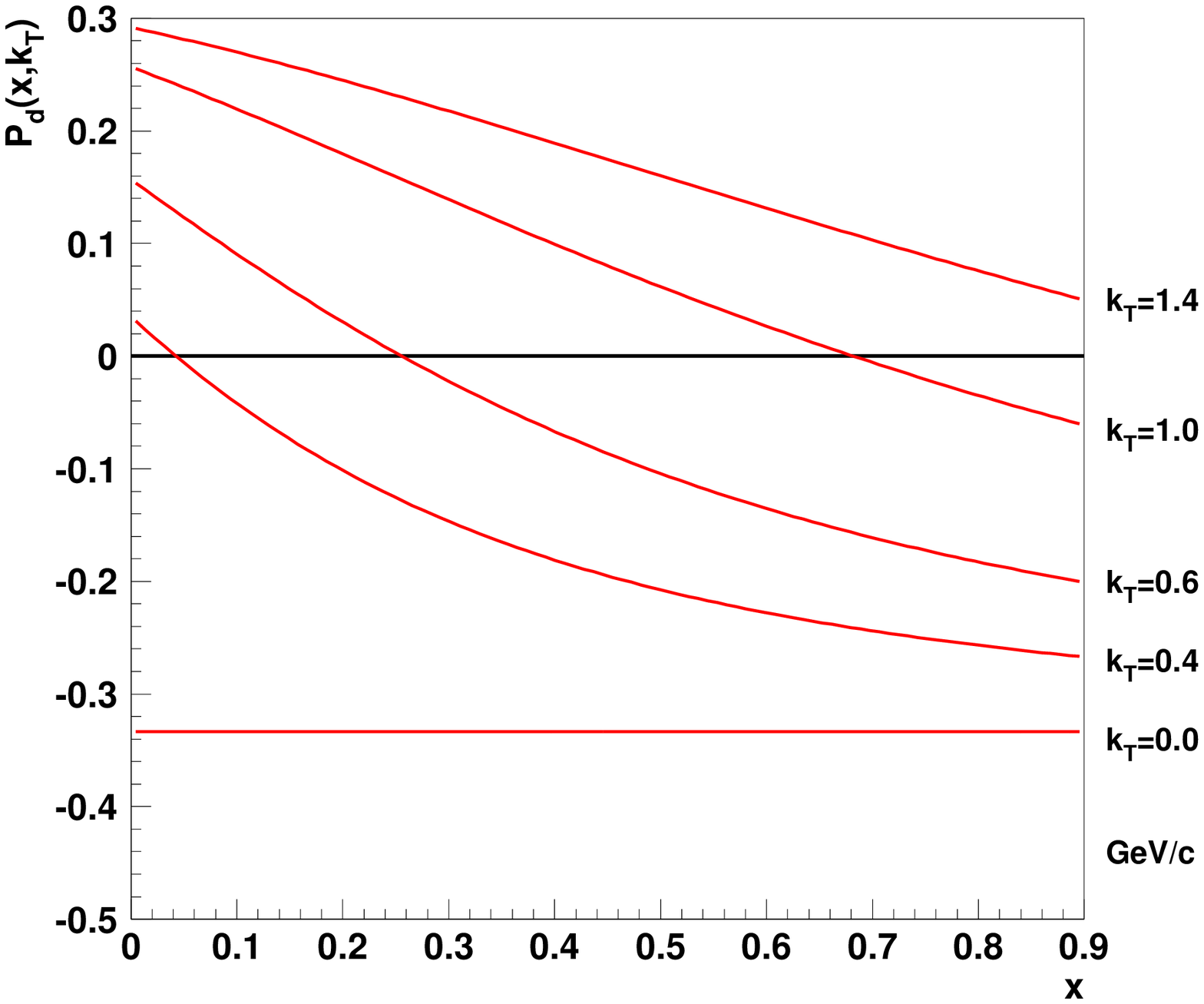,width=5.6cm}
\end{center}
\vspace{-0.75cm}
\caption{The {\it u}- (left) and {\it d}-quark (right) polarization for different values of $k_T$.}
\label{fig:fig1}
\end{figure}

In contrast, with our choice of quark-diquark model, not only does the x-dependence of $P_q(x,{\bf k}_T^2)$ changes its form for different fixed $k_T$ but it changes sign  as well, see Fig.~\ref{fig:fig1}.

This behavior is tightly related to different quark orbital momentum contributions to DFs with scalar and vector diquark. Indeed, from Eqs.~(\ref{6dfs}) the positive and negative helicity DFs  $q_{\pm}^R(x,{\bf k}_T^2)$ look as

\begin{eqnarray}\label{qpm}
    q_{+}^S(x,{\bf k}_T^2) &=& f_0^S(x)(xM+m)^2G({\bf k}_T^2,x), \nonumber \\
    q_{-}^S(x,{\bf k}_T^2) &=& f_0^S(x){\bf k}_T^2G({\bf k}_T^2,x), \nonumber \\
    q_{+}^A(x,{\bf k}_T^2)& =& {1 \over 3}f_0^A(x)\left(2(xM+m)^2+{\bf k}_T^2\right)G({\bf k}_T^2,x),\\
    q_{-}^A(x,{\bf k}_T^2) &=& {1 \over 3}f_0^A(x)\left((xM+m)^2+2{\bf k}_T^2\right)G({\bf k}_T^2,x). \nonumber
\end{eqnarray}
 The DF $q_{+}^S$ has the relative angular momentum  $L=0$ and for $q_{-}^S$ we have $L=1, \;l_z=1$. For the axial-vector diquark case $L=0$ and $L=1$ orbital momenta contribute both in $q_{+}^A$ and $q_{-}^A$ (for more discussion, see~\cite{BHMS,HA07}).

 The $x$-$k_T$ factorization assumption was used in some analyzes of high-$p_T$ hadron production asymmetries where scattering on quark is considered as a background process in polarized gluon distribution extraction. It is clear that the effective range of intrinsic $k_T$ depends on the lower cut in hadron $p_T$ and one can conclude from Fig.~\ref{fig:fig1} that the magnitude of this background contribution can be different depending on the choice of $k_T$-dependence of TMD DFs. This consideration demonstrates that it is very important to obtain a reliable information on unpolarized and helicity TMD DFs by precise measurement of $x$-, $z$-,  $p_T$- and azimuthal dependences of unpolarized SIDIS cross section and $A_{LL}$ asymmetry and perform a flavor analysis (see also~\cite{AEKP}).

\section{Azimuthal asymmetries on transversely polarized target}\label{sec:tr-asym}

Let us start by asymmetries which are described in the parton model by convolutions of twist-two DFs and FFs. The Sivers function is equal to zero in quark-diquark model, thus $A_{UT}^{\sin(\phi_h-\phi_S)}=0$\footnote{For notations and definition of azimuthal asymmetries see~\cite{AK07, BP07}}\/.

\begin{figure}
\begin{center}
\psfig{file=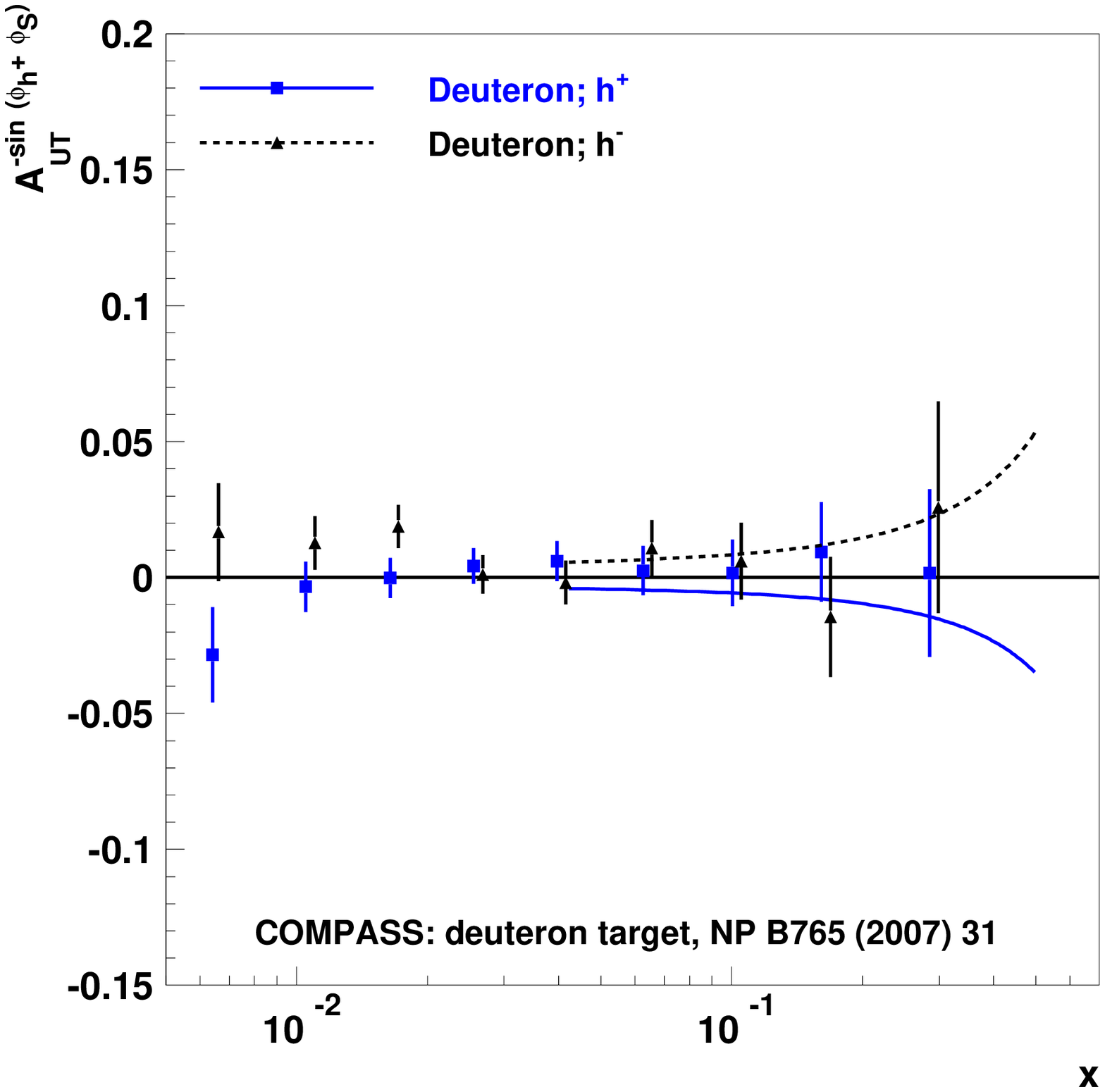,width=5.6cm}
\psfig{file=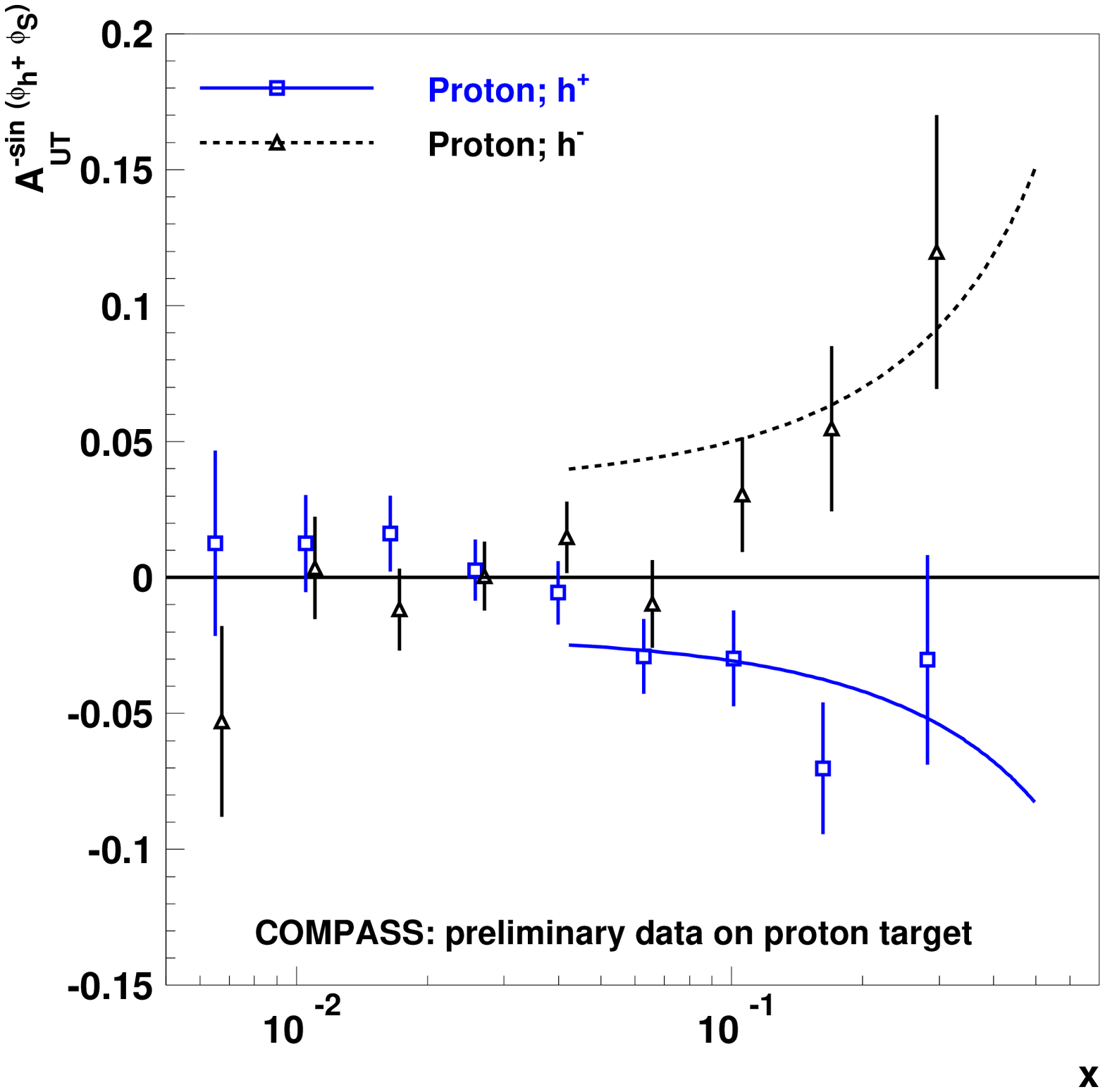,width=5.6cm}
\end{center}
\vspace{-0.75cm}
\caption{Collins asymmetry for charged hadron production.}
\label{fig:collins}
\end{figure}

The Collins asymmetry can be expressed as
\begin{equation}\label{eq:collins}
    A_{UT}^{\sin (\phi _h +\phi _S )} \propto  \frac{h_1^q \otimes H_{1q}^{\bot h}}
    {f_1^q \otimes D_{1q}^{h}},
\end{equation}
where $\otimes$ denotes integration over intrinsic transverse momentum of quark and hadron ${\it z}$- and ${\it p_T}$-variables over the kinematic domain of experiment. For details see Refs.~\cite{AK95, B06, Ans07}\/.
The results of calculations for charged hadron production asymmetry on deuteron~\cite{comp1} and proton~\cite{comp2} target are presented in left and right panels of Fig.~\ref{fig:collins}. For unpolarized FFs, we are using the recent DSS LO set~\cite{DSS} and for Collins FFs the parametrization from the recent global analysis~\cite{Prok08, Ans07}.

\begin{figure}
\begin{center}
\psfig{file=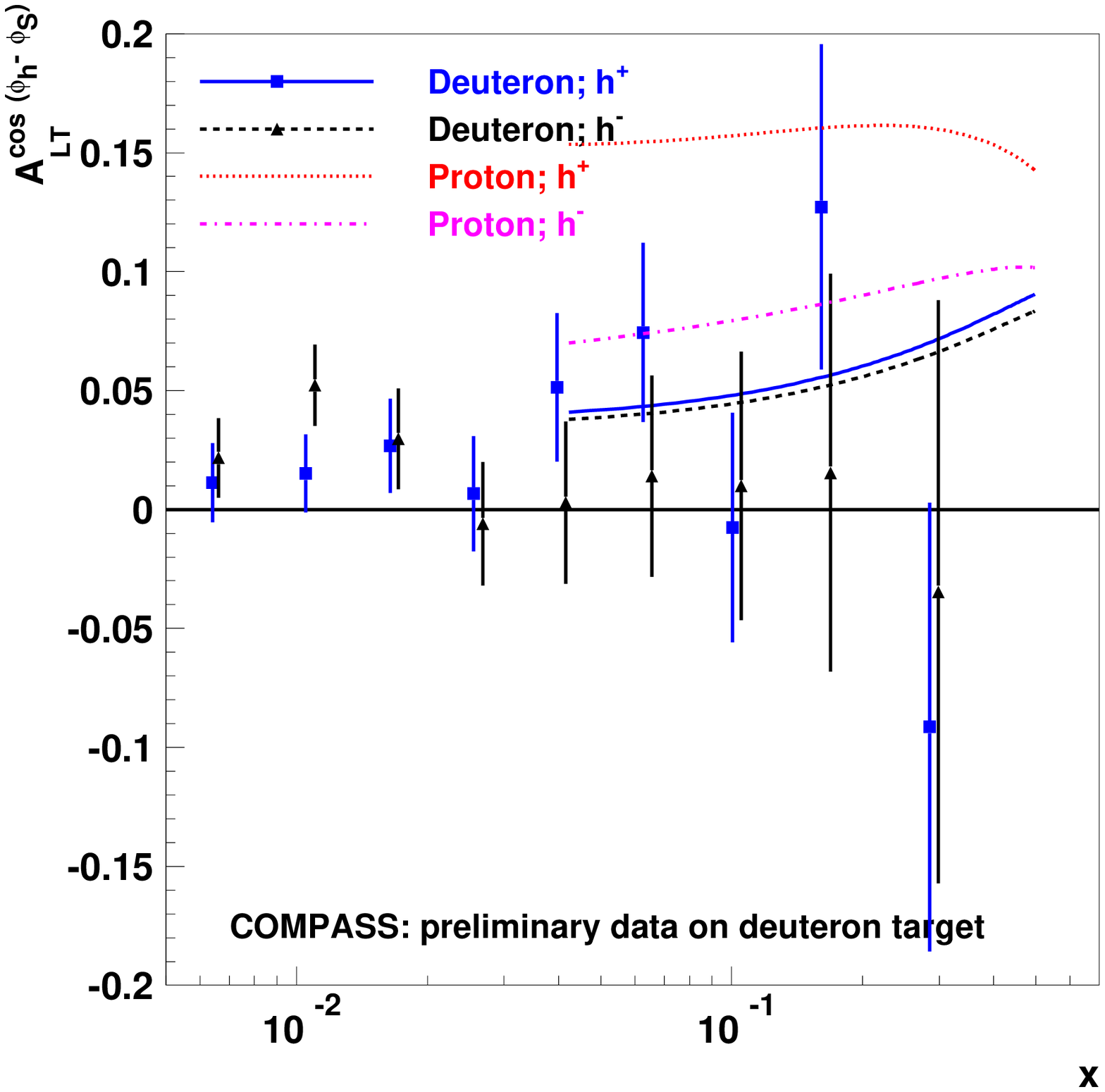,width=5.6cm}
\psfig{file=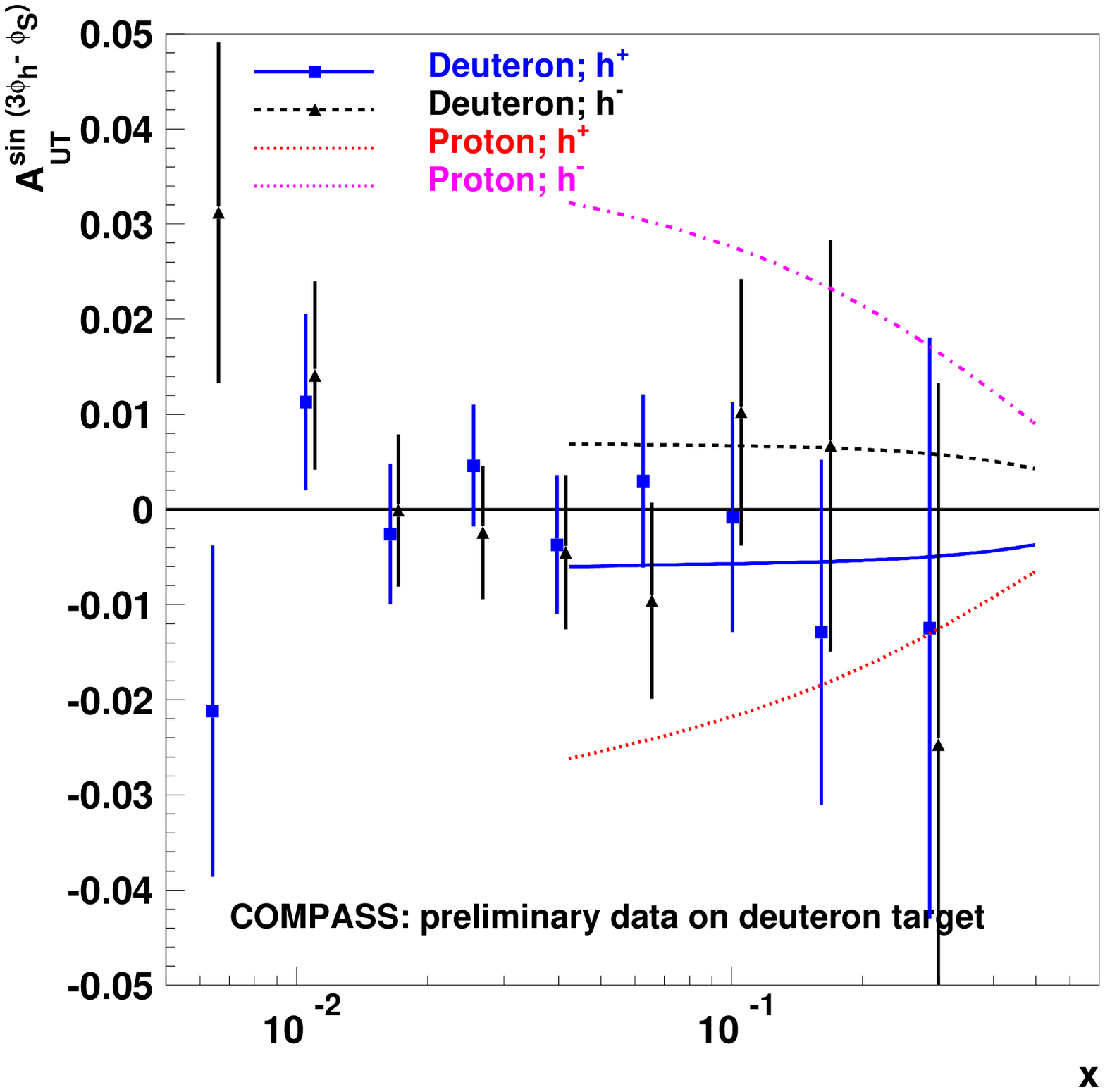,width=5.6cm}
\end{center}
\vspace{-0.75cm}
\caption{The $A_{LT}^{\cos (\phi _h -\phi _S )}$ (left) and $A_{UT}^{\sin (3\phi _h -\phi _S )}$ (right) asymmetries.}
\label{fig:fig3}
\end{figure}

The DF $g_{1T}^\perp$ gives origin to double spin asymmetry
\begin{equation}\label{eq:g1t}
    A_{LT}^{\cos (\phi _h -\phi _S )} \propto  \frac{g_{1T}^{\perp\,q} \otimes D_{1q}^{h}}
    {f_1^q \otimes D_{1q}^{h}}
\end{equation}
and the DF $h_{1T}^\perp$ --- to
\begin{equation}\label{eq:h1tp}
    A_{UT}^{\sin (3\phi _h -\phi _S )} \propto  \frac{h_{1T}^{\perp\,q} \otimes H_{1q}^{\bot h}}
    {f_1^q \otimes D_{1q}^{h}}.
\end{equation}
The results are presented in left and right panels of Fig.~\ref{fig:fig3}.

Other four target transverse spin dependent asymmetries can be treated as kinematic ${\cal O} (1/Q)$ corrections to twist-two contributions~\cite{AK95, B06}. One has
\begin{equation}\label{eq:g1t3}
    A_{LT}^{\cos (\phi _S )} \propto {M \over Q} \frac{g_{1T}^{\perp\,q} \otimes D_{1q}^{h}}
    {f_1^q \otimes D_{1q}^{h}},
\end{equation}
\begin{equation}\label{eq:g1t33}
    A_{LT}^{\cos (2\phi_h\phi _S )} \propto {M \over Q} \frac{g_{1T}^{\perp\,q} \otimes D_{1q}^{h}}
    {f_1^q \otimes D_{1q}^{h}},
\end{equation}
\begin{equation}\label{eq:h1-siv}
    A_{UT}^{\sin (\phi _S )} \propto {M \over Q} \frac{\left(h_1^q \otimes H_{1q}^{\perp h}+f_{1T}^{\perp\,q} \otimes D_{1q}^{h}\right)}
    {f_1^q \otimes D_{1q}^{h}},
\end{equation}
\begin{equation}\label{eq:h1-f1tp}
    A_{LT}^{\sin (2\phi_h\phi _S )} \propto {M \over Q} \frac{\left(h_{1T}^{\perp\,q} \otimes H_{1q}^{\perp h}+f_{1T}^{\perp\,q} \otimes D_{1q}^{h}\right)}
    {f_1^q \otimes D_{1q}^{h}}.
\end{equation}

\begin{figure}
\begin{center}
\psfig{file=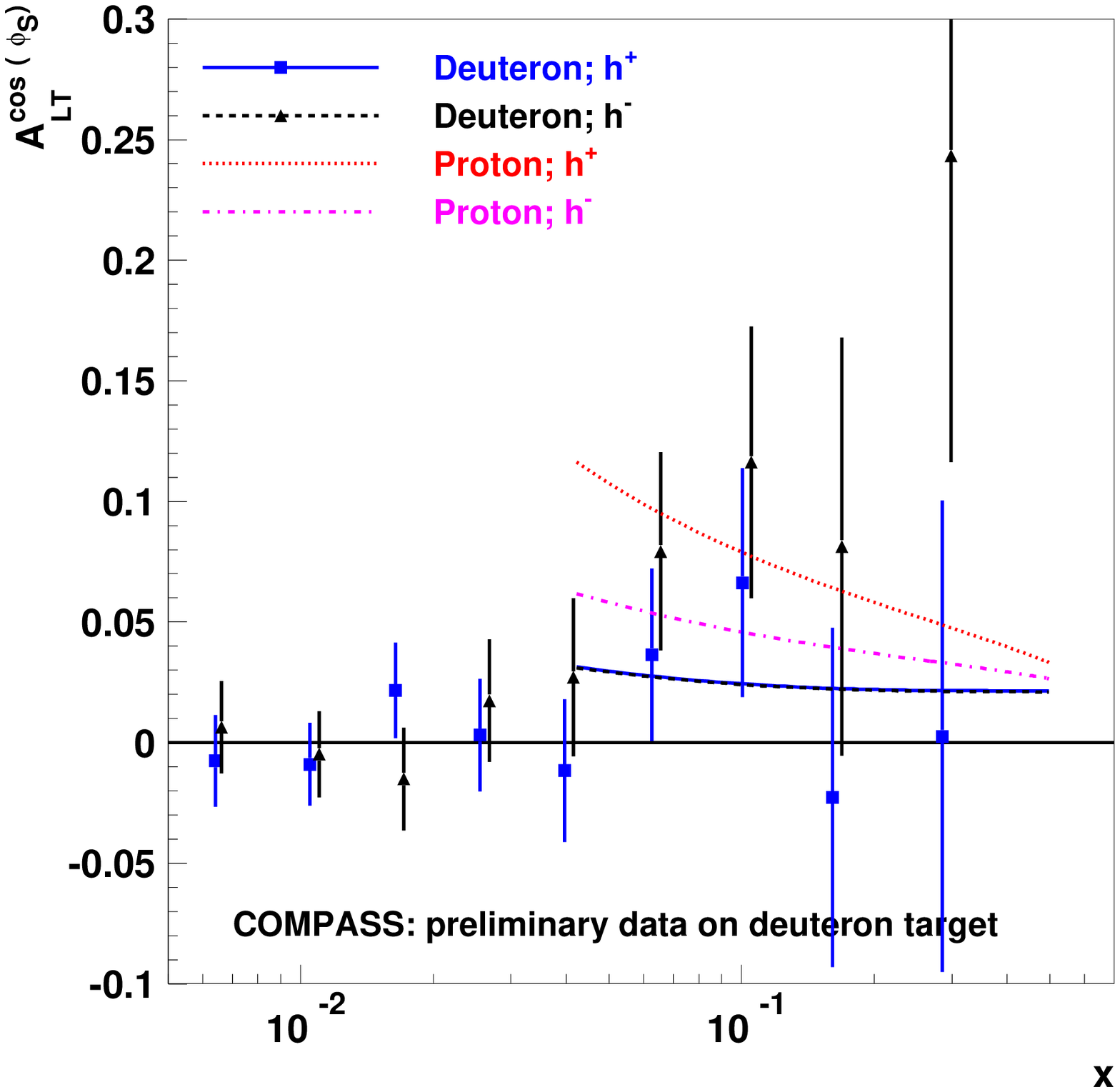,width=5.6cm}
\psfig{file=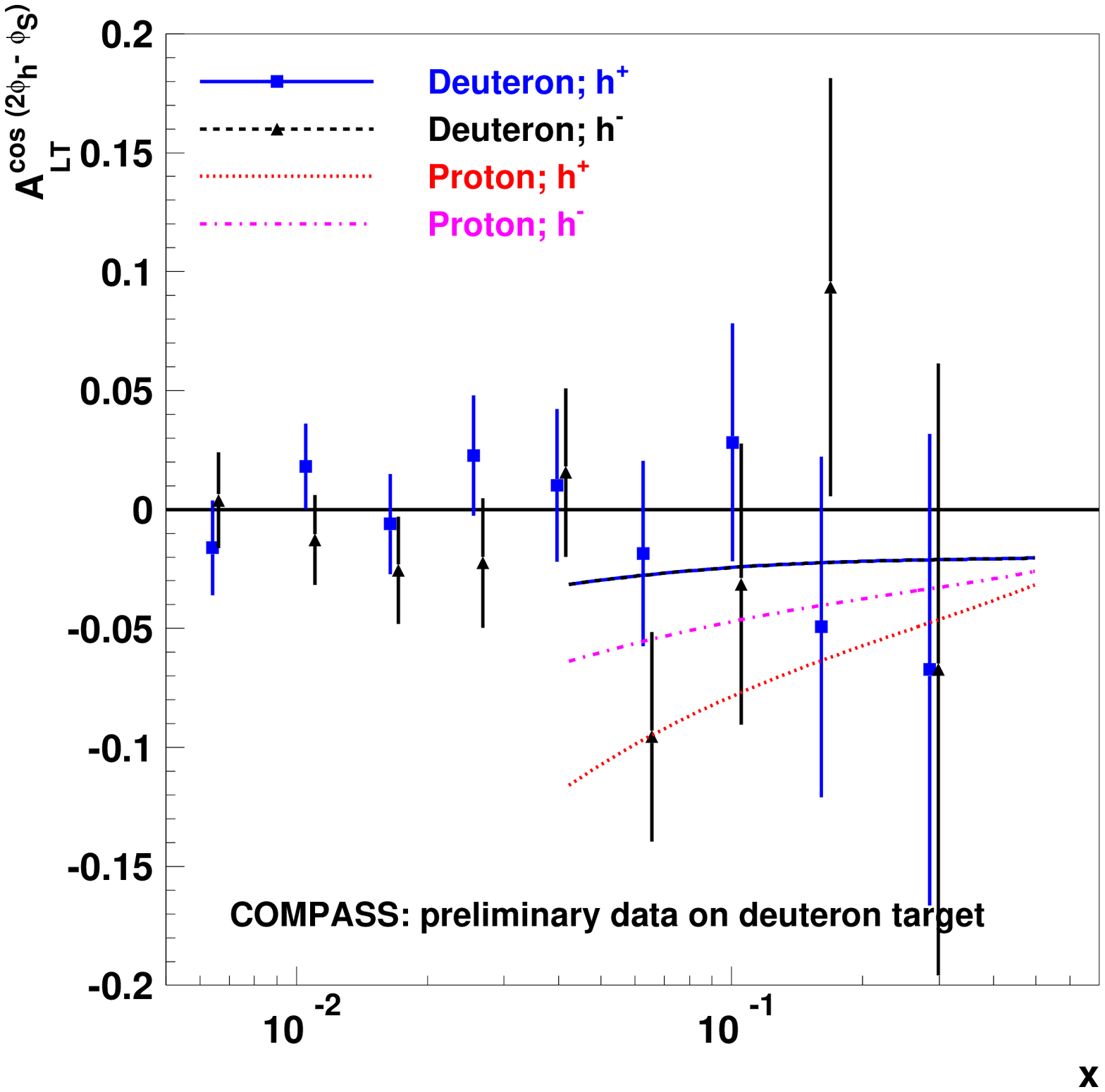,width=5.6cm}
\end{center}
\vspace{-0.75cm}
\caption{Twist-three asymmetries $A_{LT}^{\cos (\phi _S )}$ (left) and $A_{LT}^{\cos (2\phi _h -\phi _S )}$ (right).}
\label{fig:fig4}
\end{figure}

\begin{figure}
\begin{center}
\psfig{file=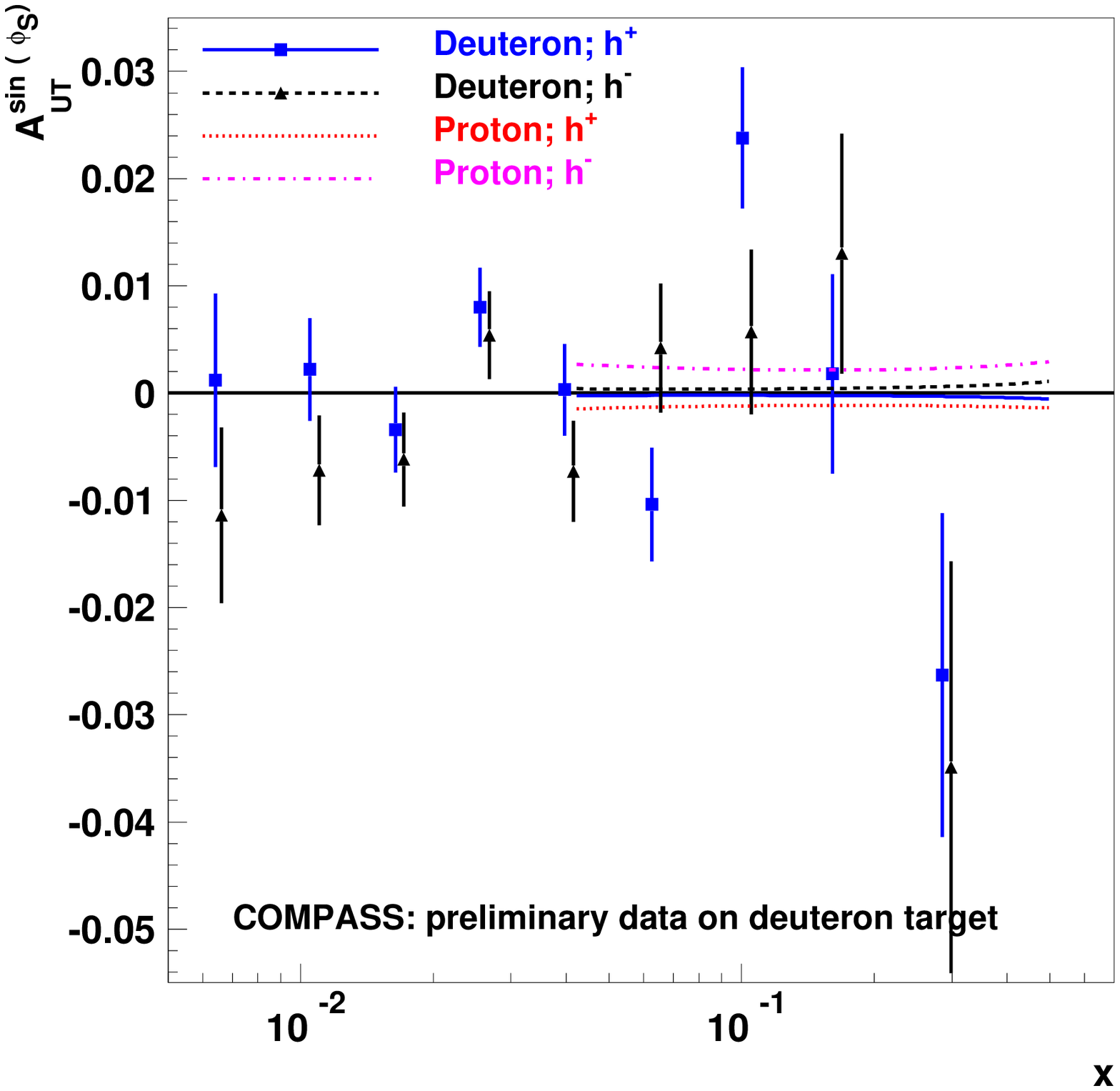,width=5.6cm}
\psfig{file=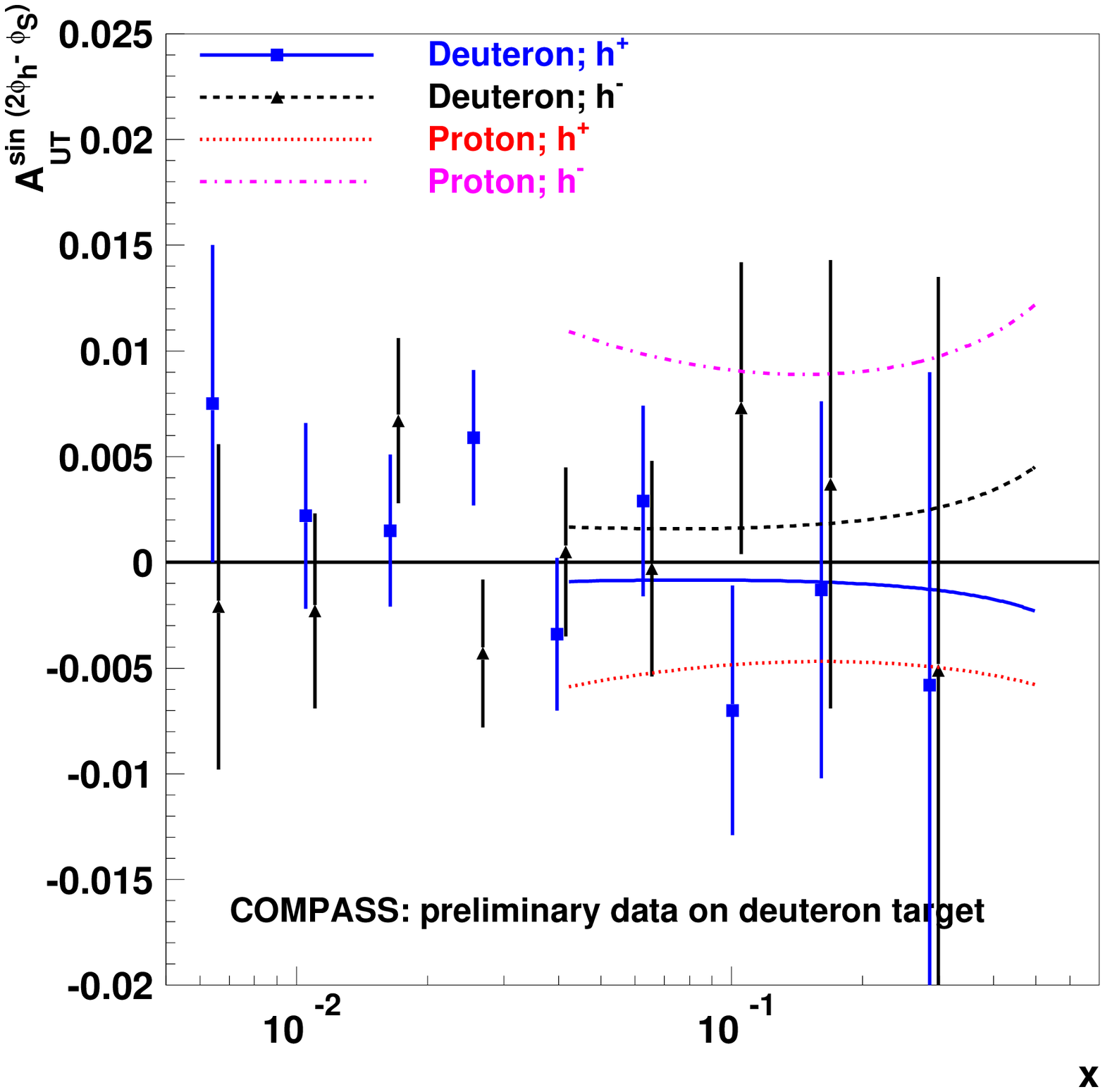,width=5.6cm}
\end{center}
\vspace{-0.75cm}
\caption{Twist-tree asymmetries $A_{UT}^{\sin (\phi _S )}$ (left) and $A_{uT}^{\sin (2\phi _h -\phi _S )}$ (right).}
\label{fig:fig5}
\end{figure}

The results are presented in Figs.~\ref{fig:fig4} and~\ref{fig:fig5}.

Note that, because the simple quark-diquark model is developed for valence quarks, the calculations are presented for not too small values of Bjorken variable {\it x}. Calculations of $z$- and $p_T$-dependences of asymmetries for COMPASS kinematics include $x$-integration starting from very small $x \approx 0.003$. This means that the main contribution to unpolarized cross-section $z$- and $p_T$-dependence comes from sea quark region. For this reason, the quark-diquark model is not well adapted to describe these dependences of asymmetries at COMPASS kinematics.

\section{Remarks on Sivers function}\label{sec:seivers}

The mechanism of final state interaction (FSI) for generating the Sivers asymmetry in SIDIS was proposed in Ref.~\cite{Brodsky:2002cx}\/. Application of this mechanism to quark-diquark model was considered in~\cite{Gamb07, Bacc03, Radi08} and~\cite {EHK}\/. With Sivers DFs calculated within this approach, one can rather well describe the $x$-dependences of pion asymmetries observed by HERMES and COMPASS. However, we have observed that these functions are violating the naive positivity constraint for Sivers analyzing power:
\begin{equation}\label{anpow}
\left | A_{Siv}(x,{\bf k}_T^2) \right | \leq 1, \qquad
A_{Siv}(x,{\bf k}_T^2)={k_T \over M} {f_{1T}^\perp(x,{\bf k}_T^2) \over f_1(x,{\bf k}_T^2)}.
\end{equation}

Namely, with dipole-like formfactor in the proton-quark-diquark vertex in models~\cite{Bacc03, Radi08, EHK} one obtains for high-$k_T$ $A_{Siv}(x,{\bf k}_T^2) \propto k_T \rightarrow \infty$, with Gaussian choice Eq. (\ref{ff}) $A_{Siv}(x,{\bf k}_T^2) \propto 1/k_T \exp\left({\bf k}_T^2/2\mu(x)^2\right) \rightarrow \infty$ and even stronger divergence in approach of~\cite{Gamb07}\/.
The similar positivity violation for polarizing power, $P_{BM}(x,{\bf k}_T^2) = \left(k_T / M \right) \left(h_{1}^\perp(x,{\bf k}_T^2) / f_1(x,{\bf k}_T^2)\right)$, holds for the Boer-Mulders function calculated in~\cite{Gamb07} --- $P_{BM}(x,{\bf k}_T^2)\rightarrow \infty$ at high $k_T$.

\section{Conclusions}

The quark-diquark model provides a good tool to study nonperturbative spin dynamics.
The predictions of target transverse spin dependent asymmetries obtained using this model are in good agreement with COMPASS measurement~\cite{AK07, BP07}.

The interesting property of this model is the violation of the $x$-$k_T$ factorization and dependence on $x$ of width of transverse momentum distribution. It is be very important to measure the SIDIS cross-sections and asymmetries with high precision and perform a global analysis of data to get reliable two-dimensional parameterizations of different TMD DFs.

Sivers and Boer-Mulders functions obtained by applying the FSI mechanism to quark-diquark model do not satisfy the positivity bounds. Therefore, the questions arise: is the FSI mechanism~\cite{Brodsky:2002cx} a universal explanation for nonzero Sivers function? Is it possible to avoid the violation of positivity by adding a formfactor in the gluon-diquark vertex or by a re-summation of higher order diagrams? More studies are needed to answer these questions.

\end{document}